\documentclass[sigconf]{acmart}

\usepackage{multirow}
\usepackage{colortbl}
\usepackage{amsmath}

\AtBeginDocument{%
  \providecommand\BibTeX{{%
    \normalfont B\kern-0.5em{\scshape i\kern-0.25em b}\kern-0.8em\TeX}}}

\begin{document}

\title{European 5G Security in the Wild: Reality versus Expectations}

\setcopyright{none}
\copyrightyear{2023}
\acmYear{2023}
\acmDOI{10.1145/xxxxxxx.xxxxxxx}
\acmConference{Conference’23}{Month}

\author{Oscar Lasierra}
\email{oscar.lasierra@i2cat.net}
\affiliation{%
  \institution{i2CAT Foundation}
  \country{}
}

\author{Gines Garcia-Aviles}
\email{gines.garcia@i2cat.net}
\affiliation{%
  \institution{i2CAT Foundation}
  \country{}
}
\author{Esteban Municio}
\email{esteban.municio@i2cat.net}
\affiliation{%
  \institution{i2CAT Foundation}
  \country{}
}

\author{Antonio Skarmeta}
\email{skarmeta@um.es}
\affiliation{%
  \institution{University of Murcia}
  \country{}
}

\author{Xavier Costa-Pérez}
\email{xavier.costa@i2cat.net}
\affiliation{%
    \institution{i2CAT Foundation and ICREA}
    \institution{NEC Laboratories Europe}
    \country{}
}

\renewcommand{\shortauthors}{Oscar Lasierra, Gines Garcia-Aviles, Esteban Municio, Antonio Skarmeta, \& Xavier Costa-Pérez}

\newcommand{\newtext}[1]{{\color{blue}{#1}}}
\newcommand{\olh}[1]{{\color{red}{[OLH]: #1}}}
\newcommand{\emh}[1]{\textcolor{olive}{[EMH] #1}}
\newcommand{\gga}[1]{{\color{orange}{(GGA): #1}}}
\definecolor{grOK}{RGB}{34,139,34}
\definecolor{redN}{RGB}{139,0,0}
\definecolor{orgR}{RGB}{255,165,0}
\definecolor{bubblegum}{rgb}{0.99, 0.76, 0.8}
\definecolor{darkseagreen}{rgb}{0.56, 0.74, 0.56}
\definecolor{flax}{rgb}{0.93, 0.86, 0.51}

\definecolor{Optional5G}{RGB}{158,154,200}
\definecolor{Mandatory5G}{RGB}{55,126,184}
\definecolor{In5gStandard}{RGB}{77,175,74}
\definecolor{In4gStandard}{RGB}{152,78,163}
\definecolor{NoSecurity}{RGB}{228,26,28}
\definecolor{NotImplemented}{RGB}{255,127,0}
\definecolor{NoData}{RGB}{166,86,40}

\begin{abstract}
5G cellular systems are slowly being deployed worldwide delivering 
the promised unprecedented levels of throughput and latency to hundreds of millions of users. At such scale security is crucial, and consequently, the 5G standard includes a new series of features to improve the security of its predecessors (i.e., 3G and 4G). 
In this work, we evaluate the actual deployment in practice of the promised 5G security features by analysing current commercial 5G networks from several European operators. 
By collecting 5G signalling traffic in the wild in several cities in Spain, we i) fact-check which 5G security enhancements are actually implemented in current deployments, ii)  provide a rich overview of the implementation status of each 5G security feature in a wide range of 5G commercial networks in Europe and compare it with previous results in China, iii) analyse the implications of optional features not being deployed, and iv) discuss on the still remaining 4G-inherited vulnerabilities. 
Our results show that in European 5G commercial networks, the deployment of the 5G security features is still on the works. This is well aligned with results previously reported from China~\cite{nie2022measuring} and keeps these networks vulnerable to some 4G attacks, during their migration period from 4G to 5G.

\end{abstract}



\settopmatter{printacmref=false}
\settopmatter{printfolios=true}
\maketitle

\section{Introduction}
\label{sec:intro}

The arrival of the fifth generation of mobile networks (5G) is substantially changing the way networks are designed and deployed. From the subscribers perspective, 5G effectively provides an improved performance compared with their predecessors, increasing available bandwidth (e.g., to provide on-demand high-quality video services) and reducing end-to-end latency (e.g., to provide real-time augmented/virtual reality applications). By the end of 2021, more than 176 commercial 5G networks have been deployed worldwide, of which only 22 were already 5G Stand Alone (SA) networks. 
~\cite{gsma2022}. Unfortunately, such growing figures also bring greater risks in terms of security.

However, unlike previous mobile generations such as 3G/4G which are subject to a number of known attacks~\cite{mjolsnes2017easy,kohls2019lost,shaik2015practical,rupprecht2020imp4gt}, 5G provides security enhancements through a series of new generation specifications defined by the 3rd Generation Partnership Project (3GPP), including TS~33.501~\cite{ts33501} and TS~33.511~\cite{ts33511}. Despite this, while current real-world 5G deployments follow the same architectural security framework reference, neither all of them implement the same 5G security mechanisms enabled by the new specifications, nor they do it in the same way. This is usually caused by the optionality of some mechanisms and by the operators' inherent constraints (cost, compatibility, or performance)~\cite{nie2022measuring}.

In this work, we report a hands-on security analysis of currently deployed 5G networks, fact-checking security mechanisms compliance and identifying still existing vulnerabilities in current 5G deployments.
For those non-compliant or partially-compliant deployments, we identify and provide an in-depth characterisation of the attacks they are vulnerable to.

In order to perform such analysis, we collect and study signalling messages between various 5G networks and the User Equipment (UE) through commercial cellular traffic sniffers focusing on currently deployed 5G networks from different network operators in urban and suburban areas of various cities in the east coast of Spain.
These traces include information about the User Plane (UP) and Control Plane (CP) security activation, the subscriber identifiers exchanged and the Authentication procedures performed for accessing the network. 

Our measurements show that although commercial deployments do not implement all user authentication mechanisms specified in the standard, the confidentiality and integrity implementation at the UE does always seem to comply with the standard. However, unlike previously reported in ~\cite{nie2022measuring} for Chinese 5G deployments in Beijing, the majority of the observed networks are still exposed to 4G-inherited vulnerabilities such as identity and user data leakage and Denial of Service (DoS) attacks because of the yet general absence of Standalone (SA) 5G 
network deployments. Note that this is as expected due to practical deployment reasons; which is aligned with the roadmap specified by operators towards the adoption of 5G not just in Europe but worldwide. GSMA forecasts a 44\% average adoption of 5G within Europe by 2025~\cite{GSMA-5G}. So the migration path from 4G to 5G is on the works but will take some years still to be completed.

Therefore, the main contributions of this work are i) a comprehensive compliance analysis for different 5G networks deployed in Spain in order to fact-check and evaluate the actual security and privacy mechanisms implemented by vendors and operators in a typical European 5G network deployment\footnote{The network operators considered in this work operate in about 70\% of the countries in the EU with similar 5G deployments}; and ii) a study of the available security vulnerabilities in current commercial 5G networks.

The rest of this work is structured as follows. In Section~\ref{sec:background} we briefly provide the necessary background on 5G NR. Section~\ref{sec:security_in_5gnr} describes the corresponding security mechanisms included in the 5G standard and identifies the most common security threats. 
In Section~\ref{sec:metrics_in_the_wild} we detail the methodology followed for data collection and its subsequent analysis
and in Section~\ref{sec:evaluation} we report the results,
extensively discussing the capabilities, standard compliance, and vulnerabilities observed in the different 5G networks. Finally, Section~\ref{sec:conclusions} concludes this work.

\section{Background}
\label{sec:background}

\subsection{5G Outline}

The architecture of 5G cellular networks can be logically separated into three main components, User Equipment (UE), the Radio Access Network (RAN), and the mobile Core Network (CN).
The UEs establish a wireless connection with the RAN to be able to reach the CN, which acts as i) an authentication entity, allowing/denying devices to access the network; and ii) acting as an ingress/egress point of the traffic generated from/to the internet.

Within the 5G context, UEs are essentially defined as a combination of two components.
First, the Universal Subscriber Identity Module (USIM) card, which is used to store user identification data, such as the public/private keys and the Subscriber Permanent Identifier (SUPI), known as the International Mobile Subscriber Identity (IMSI) in 4G. Second, the Mobile Equipment (ME) hardware itself, is identified by the International Mobile Equipment Identity (IMEI).
     
The RAN manages the wireless connectivity through the 5G base stations (gNBs), replacing or coexisting with legacy 4G base stations (eNBs). LTE/NR coexistence is ensured through the 5G Non-standalone (NSA) mode or EUTRA NR Dual Connectivity (ENDC), which allows UEs to configure a 5G secondary node for data plane transmissions. This mode keeps 4G eNBs as master nodes which are in charge of carrying control plane traffic. 
In contrast, 5G Standalone (SA) mode adopts the gNB as the master node of the connection to jointly manage both data and control planes traffic.
The interaction between the UE and the RAN is one of the most vulnerable parts in the network, and therefore, the main security features imposed by the 5G standard come to solve some of the major risks and pitfalls in the wireless domain~\cite{cheng2022new,shaik2015practical,palama2021imsi, cui2022security}.
     
Similarly to 4G, the 5G core network (CN) provides the UEs with external packet data network connectivity. 
It consists of various network functions to manage different 
fundamental processes such as session control (SMF), authentication (AUSF and SEAF), access and mobility (AMF), etc.   

\subsection{Critical NR Procedures: Initial Attachment and Registration}
\label{subsec:nr_procedures}
The initial procedures performed by 5G NR carry essential information required to establish a stable and secure communication through the RAN. These processes are based on the exchange of information between parties: \textit{UE-gNB} for the radio link and \textit{UE-gNB-CN} for a higher-level communication layer. 
Both processes must be performed in a way that preserves security and confidentiality, avoiding third-party observers to gather the exchanged information, and hence bypassing security leaks in subsequent communications. 

\begin{figure}
\centering
\includegraphics[width=0.45\textwidth]{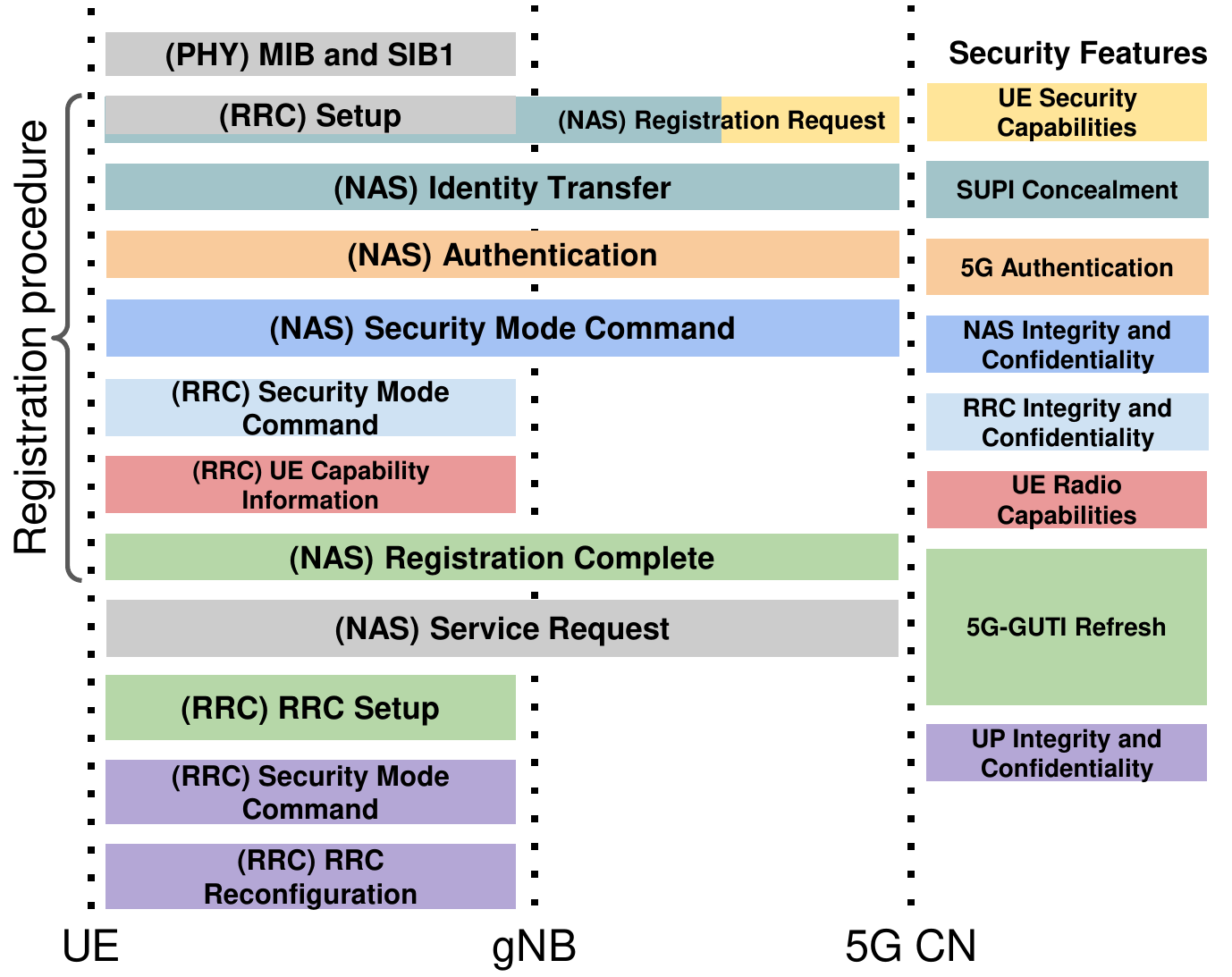}
\caption{5G NR Initial Registration Procedure}
\label{fig:sa_registration}
\vspace{-6mm}
\end{figure}

\subsubsection{\textbf{Broadcast Channel and Random Access:}}

To effectively establish a connection, the UE must perform a set of interactions with the gNB before starting with the registration procedure itself, the Cell search and the Random Access Channel (RACH). 

The Cell Search procedure allows the UE to acquire time and frequency synchronisation within cells with the goal of retrieving cell parameters and system information from the Master Information Block (MIB) and the System Information Block (SIB). Synchronisation is obtained by detecting Synchronisation Signal Block (SSB) and decoding Primary and Secondary Sinchronisation Signals (PSS and SSS) located on the synchronisation raster. Then, MIB is decoded from the Physical Broadcast Channel (PBCH) from the same synchronisation raster to subsequently configure the Control resource set zero (CORESET0) and SearchSpace. With this information, the UE can perform the blind decode of the Physical Downlink Control Channel (PDCCH) and configure the remaining parameters to find and decode SIB1 in the Physical Downlink Shared Channel (PDSCH) (full procedure defined in 3GPP 38.104~\cite{ts38104}).

RACH procedure allows the UE to configure UL synchronisation and obtain an identifier for the radio communication. If BeamForming is supported, the UE shall detect, choose and synchronise with the best beam to start the communication with the gNB.

\subsubsection{\textbf{Radio Resource Control:}}
After the Random Access procedure, if the UE is not attached to the network, it has to initiate the registration procedure. Otherwise, 
the UE initiates the tracking area update if it has changed since the last update. For initiating any NAS procedure, the UE needs to establish a Radio Resource Control (RRC) connection with the gNB.
The main purpose of this procedure is to establish an active connection with the gNB, enabling the acquisition of radio resources for the communication.
RRC connection establishment involves the creation of the Signalling Radio Bearer one (SRB1) for the RRC messages exchange. The last message of this process can carry the initial Non-Access Stratum (NAS) message from the UE to the Access and Mobility Management Function (AMF) via the gNB (Mobility Management Entity (MME) via the eNB for NSA deployments).

\subsubsection{\textbf{Non-Access-Stratum:}}
To get Non-Access Stratum (NAS)-level services (e.g,
internet connectivity)
, NAS nodes in the network need to know about the UE. To facilitate this, the UE has to initiate the Attach Procedure, which is mandatory to be performed by the UE at boot time (or by setting Airplane mode off). 
Once the attach procedure succeeds, a context is established for the UE in the core,
and a default bearer is established between the UE and the Packet Data Network Gateway (PDN GW). This results in the allocation of an available IP address to the UE, enabling IP-based internet services in the 5G device.

\section{Security in 5G NR}
\label{sec:security_in_5gnr}

Security in cellular networks has been evolving during the different mobile generations in order to address the open threads identified during their operation. The enhancements brought by the 5G standard~\cite{ts33501} are depicted in Figure~\ref{fig:sa_registration} and summarised next.

\subsection{UE credentials and identifiers}
\label{subsec:5gnr_ids}

One of the major enhancements introduced in 5G SA networks is the concealment of the SUPI. In previous generations, subscriber permanent
identifiers (which in some cases contain relevant information such as the phone number) were sent in clear text and thus, attackers could retrieve this information and perform impersonation attacks~\cite{rupprecht2020imp4gt,mjolsnes2017easy}. In 5G, the UE sends a concealed version of the SUPI
called Subscription Concealed Identifier (SUCI)
generated by using asymmetric cryptography (the private key is securely stored in the USIM). However, 5G SUCI-Catching attacks are still possible as reported in~\cite{chlosta20215g}. In this sense, in order to avoid sending the subscriber identifier over the radio link, temporal identifiers were added in 4G (Globally Unique Temporary Identity, GUTI and the Temporary Mobile Subscriber Identity, TMSI) but their refresh rate was sub-optimal, failing on providing confidentiality and anonymity to users. 5G networks also use temporary identifiers but 3GPP imposes specific guidelines dealing with aforementioned vulnerabilities.

\subsection{Enhanced Authentication and Privacy}
\label{subsec:5gnr_auth}

The 5G Authentication and Key Agreement (5G-AKA) protocol is a security protocol introduced in the 5G standard to provide mutual authentication between all the components in the communication, i.e., UE, Serving Network (SN) and Home Network (HN) with privacy-preserving policies (i.e., providing user ID confidentiality and preventing from user tracking). Similarly to AKA protocols of previous generations, 5G-AKA allows two end-points to establish the root keys from which new keys in subsequent security procedures will be derived~\cite{ROMMER2020171}. However previous authentication protocols, such as 4G-AKA, failed to provide 
anonymity to the user because a) the IMSI of the user was sent in plain text and b) when replacing the ID of the user with temporary identifiers they are usually static and persistent, hence predictable as studied in~\cite{hong2018guti}. 5G-AKA adopts the use of the 5G-GUTI to address this issue. Additionally, 5G-AKA enables Non-3GPP accesses (authentication is no longer related to one specific access technology) and allows the Serving Network and the Home Network to mutually authenticate themselves by cross-verification
(i.e., by the AMF and SEAF in the Serving Network and by the AUSF in the Home Network)~\cite{koutsos20195g}.
Table~\ref{table:5gnr_auth_aka} summarizes the 5G-AKA security enhancements along with the new 5G UE identifiers, in contrast with the previous 4G-AKA protocol.

\begin{table}[]
\scriptsize
\begin{tabular}{cc|c|c}
 & & 4G-AKA & 5G-AKA \\ \hline
  & UE & USIM& USIM   \\
 Confidentiality & SN & MME & AMF/SEAF  \\ 
 & HN & HSS & AUSF \\\hline
{UE Identity}  &  & IMSI/GUTI  & SUCI/5G-GUTI         \\  
{Trust Model} &  & Shared Symmetric Key & Shared Symmetric Key \\ 
{UE Authentication} &  & No information to HN & Inform HN  \\ 
\end{tabular}
\caption{5G-AKA Security Enhancements}
\label{table:5gnr_auth_aka}
\vspace{-10mm}
\end{table}

\subsection{Improved Confidentiality and Integrity}
\label{subsec:5gnr_confidentiality}

Previous generations of cellular networks failed on providing confidentiality/integrity protection on some pre-authentication signalling messages, allowing attackers to 
exploit multiple vulnerabilities~\cite{sec4G_BreakingLTE_L2}. For that reason, 5G introduces novel protection mechanisms specifically designed for signalling and user data.
Besides increasing the length of the key algorithms (to 256-bit expected for future 3GPP releases), 5G forces mandatory integrity support of the user plane, and extends confidentiality and integrity protection to the initial NAS messages. Table~\ref{table:reality_check_results} summarises in column \textit{Standard} the requirements in terms of confidentiality and integrity protection as defined in~\cite{ts33501}.

5G also secures the UE network capabilities, a field within the initial NAS message, which is used to allow UEs to report to the AMF about the supported integrity and encryption algorithms in the initial NAS message. 

In addition to backward compatibility, 5G UEs shall implement New Radio Encryption Algorithm (NEA) 0, 128-NEA1 and 128-NEA2 for confidentiality protection and New Radio Integrity Algorithm (NIA) 0, 128-NIA1 and 128-NIA2 for integrity protection. However, the implementation of the 128-NEA3 and 128-NIA3 is optional~\cite{ts33501}. In 4G, the UE security capabilities are exchanged with integrity protection only when the UE has already established a security context.  
An attacker entity could capture this message and gain substantial information, e.g.,  
the technologies supported by the UE or the device model in the best-case scenario.
In order to prevent this, 5G includes both integrity and confidentiality protection in the initial registration NAS message to protect the \textit{UE capability} field. However, for both 4G and 5G, if the UE does not have an established security context (i.e., the first registration attempt), the \textit{UE capability} field is sent in clear. This information allows an attacker to read/modify the exchanged information and perform multiple attacks (e.g., user identification and power drain ~\cite{shaik2019new}.

\subsection{UE Radio Capabilities transfer}
\label{subsec:5gnr_mandatory_mechanisms}

Before establishing the connection, the UE needs to provide the gNB its capabilities for radio access (e.g., supported frequency bands, EN-DC support, etc.). In previous generations, this information was sent without establishing CP security directives, and hence, an adversary could hijack this information and perform bidding down attacks ~\cite{shaik2019new}. 5G ensures its protection by sending them in the \textit{RRC UE Capability Information} message after enabling security directives.

\section{Metrics in the Wild}
\label{sec:metrics_in_the_wild}

\subsection{Data Collection Methodology}
\label{subsec:data_collection_methodoloty}

In order to characterise 5G commercial deployments we have used a commercial protocol analyser with two different SIM cards from two different network operators\footnote{For anonymity reasons, we will refer to them as Operator A and Operator B.}. The commercial protocol analyser is a \textit{Keysight NEMO Handy Handheld}\footnote{https://www.keysight.com/us/en/product/NTH00000B/nemo-handy-handheld-measurement-solution.html} which includes a debugging tool used for wireless diagnostics. We have collected data traces from six Spanish cities: Barcelona (B), Tarragona (T), Castellón de
la Plana (C), Valencia (V), Alicante (A) and Murcia (M) (see Table~\ref{tab:cities}).

\begin{table}[]
\scriptsize
\begin{tabular}{|l|l|l|l|l|l|l|}
\hline
\textbf{City}       & Murcia  & Alicante & Valencia & Castellon & Tarragona & Barcelona \\ \hline
\textbf{Population} & 450.000 & 330.000  & 790.000  & 170.000   & 170.000   & 1.62M     \\ \hline
\end{tabular}
\caption{Cities covered for data collection}
\label{tab:cities}
\vspace{-10mm}
\end{table}

Then, to homogenise the data collection process, we have defined an experimentation methodology consisting on the following steps:
\begin{enumerate}
    \item \textbf{Airplane mode ON}: The terminal will always start with airplane mode activated.
    \item \textbf{Start data collection}: Once the airplane mode of the terminal is active, we start the data collection tool at the device.
    \item \textbf{Airplane mode OFF}: Disabling the airplane mode will allow the device to initiate the registration process to establish an active session with the mobile operator.
    \item \textbf{Initial registration}: At this phase, we wait until the registration process is complete.
    \item \textbf{Traffic generation}: This phase consists on the generation of ICMP traffic to check the connectivity status and to force a possible reconfiguration of the radio channel.
    \item \textbf{Stop data collection}: Finally, we stop the data collection tool as well as the collected data of the experiment.
\end{enumerate}

Finally, to effectively study the temporary identifiers, we replace the \textit{Traffic Generation} step with \textit{ON-OFF Switch}, where airplane mode is activated and deactivated during the traffic gathering.
Both types of experiments were performed on each geographical place with an average duration of 15 minutes. It is important to highlight the non-intrusive nature of the data collection process, where we only collect data transmitted openly over the air in a \emph{passive} manner, i.e. without performing any interaction with the network or users.

\begin{table*}[]
\resizebox{\textwidth}{!}{
\begin{tabular}{|lll|l|l|l|llllllllllll|ccc|}
\cline{1-3} \cline{5-5} \cline{7-21}
\multicolumn{3}{|l|}{Source}                                                                                                                                                             &                     & \multicolumn{1}{c|}{Standard}                 &                     & \multicolumn{12}{c|}{Commercial}                                                                                                                                                                                                                                                                                                                                                                                                                                                                                                                                         & \multicolumn{3}{c|}{{}\cite{nie2022measuring}{}}                                                                                                            \\ \cline{1-3} \cline{5-5} \cline{7-21} 
\multicolumn{3}{|l|}{Operator}                                                                                                                                                           &                     & \multicolumn{1}{c|}{\cellcolor[HTML]{C0C0C0}} &                     & \multicolumn{6}{c|}{Operator A}                                                                                                                                                                                                                                                    & \multicolumn{6}{c|}{Operator B}                                                                                                                                                                                                                                 & \multicolumn{1}{c|}{C}                             & \multicolumn{1}{c|}{D}                             & E                             \\ \cline{1-3} \cline{5-5} \cline{7-21} 
\multicolumn{3}{|l|}{Location}                                                                                                                                                           &                     & \cellcolor[HTML]{C0C0C0}                      &                     & \multicolumn{1}{l|}{M}                        & \multicolumn{1}{c|}{A}                        & \multicolumn{1}{c|}{V}                        & \multicolumn{1}{c|}{C}                        & \multicolumn{1}{c|}{T}                        & \multicolumn{1}{c|}{B}                        & \multicolumn{1}{c|}{M}                        & \multicolumn{1}{c|}{A}                        & \multicolumn{1}{l|}{V}                        & \multicolumn{1}{c|}{C}                        & \multicolumn{1}{c|}{T}                        & \multicolumn{1}{c|}{B}   & \multicolumn{3}{c|}{Beijing}                                                                                                            \\ \cline{1-3} \cline{5-5} \cline{7-21} 
\multicolumn{1}{|l|}{}                                                                                        & \multicolumn{2}{l|}{5G AKA}                                              &                     & \cellcolor[HTML]{377EB8}                      &                     & \multicolumn{1}{l|}{\cellcolor[HTML]{E41A1C}} & \multicolumn{1}{l|}{\cellcolor[HTML]{E41A1C}} & \multicolumn{1}{l|}{\cellcolor[HTML]{E41A1C}} & \multicolumn{1}{l|}{\cellcolor[HTML]{E41A1C}} & \multicolumn{1}{l|}{\cellcolor[HTML]{E41A1C}} & \multicolumn{1}{l|}{\cellcolor[HTML]{E41A1C}} & \multicolumn{1}{l|}{\cellcolor[HTML]{E41A1C}} & \multicolumn{1}{l|}{\cellcolor[HTML]{E41A1C}} & \multicolumn{1}{l|}{\cellcolor[HTML]{E41A1C}} & \multicolumn{1}{l|}{\cellcolor[HTML]{E41A1C}} & \multicolumn{1}{l|}{\cellcolor[HTML]{E41A1C}} & \cellcolor[HTML]{E41A1C} & \multicolumn{1}{c|}{\cellcolor[HTML]{FFFFFF}-----} & \multicolumn{1}{c|}{\cellcolor[HTML]{FFFFFF}-----} & \cellcolor[HTML]{FFFFFF}----- \\ \cline{2-3} \cline{5-5} \cline{7-21} 
\multicolumn{1}{|l|}{}                                                                                        & \multicolumn{2}{l|}{SUCI}                                                &                     & \cellcolor[HTML]{377EB8}                      &                     & \multicolumn{1}{l|}{\cellcolor[HTML]{E41A1C}} & \multicolumn{1}{l|}{\cellcolor[HTML]{E41A1C}} & \multicolumn{1}{l|}{\cellcolor[HTML]{E41A1C}} & \multicolumn{1}{l|}{\cellcolor[HTML]{E41A1C}} & \multicolumn{1}{l|}{\cellcolor[HTML]{E41A1C}} & \multicolumn{1}{l|}{\cellcolor[HTML]{E41A1C}} & \multicolumn{1}{l|}{\cellcolor[HTML]{E41A1C}} & \multicolumn{1}{l|}{\cellcolor[HTML]{E41A1C}} & \multicolumn{1}{l|}{\cellcolor[HTML]{E41A1C}} & \multicolumn{1}{l|}{\cellcolor[HTML]{E41A1C}} & \multicolumn{1}{l|}{\cellcolor[HTML]{E41A1C}} & \cellcolor[HTML]{E41A1C} & \multicolumn{1}{c|}{\cellcolor[HTML]{E41A1C}}      & \multicolumn{1}{c|}{\cellcolor[HTML]{E41A1C}}      & \cellcolor[HTML]{E41A1C}      \\ \cline{2-3} \cline{5-5} \cline{7-21} 
\multicolumn{1}{|l|}{}                                                                                        & \multicolumn{1}{l|}{}                               & After Registration &                     & \cellcolor[HTML]{377EB8}                      &                     & \multicolumn{1}{c|}{\cellcolor[HTML]{4DAF4A}}                         & \multicolumn{1}{c|}{\cellcolor[HTML]{4DAF4A}}                         & \multicolumn{1}{c|}{\cellcolor[HTML]{4DAF4A}}                         & \multicolumn{1}{c|}{\cellcolor[HTML]{4DAF4A}}                         & \multicolumn{1}{c|}{\cellcolor[HTML]{4DAF4A}}                         & \multicolumn{1}{c|}{\cellcolor[HTML]{4DAF4A}}                         & \multicolumn{1}{c|}{\cellcolor[HTML]{4DAF4A}}                         & \multicolumn{1}{c|}{\cellcolor[HTML]{4DAF4A}}                         & \multicolumn{1}{c|}{\cellcolor[HTML]{4DAF4A}}                         & \multicolumn{1}{c|}{\cellcolor[HTML]{4DAF4A}}                         & \multicolumn{1}{c|}{\cellcolor[HTML]{4DAF4A}}                         & \multicolumn{1}{c|}{\cellcolor[HTML]{4DAF4A}}                         & \multicolumn{1}{c|}{\cellcolor[HTML]{4DAF4A}}      & \multicolumn{1}{c|}{\cellcolor[HTML]{4DAF4A}}      & \cellcolor[HTML]{4DAF4A}      \\ \cline{3-3} \cline{5-5} \cline{7-21} 
\multicolumn{1}{|l|}{\multirow{-4}{*}{User Authentication}}                                                   & \multicolumn{1}{l|}{\multirow{-2}{*}{GUTI Refresh}} & After Service Req. &                     & \cellcolor[HTML]{377EB8}                      &                     & \multicolumn{1}{c|}{\cellcolor[HTML]{E41A1C}}                         & \multicolumn{1}{c|}{\cellcolor[HTML]{E41A1C}}                        & \multicolumn{1}{c|}{\cellcolor[HTML]{E41A1C}}                         & \multicolumn{1}{c|}{\cellcolor[HTML]{E41A1C}}                         & \multicolumn{1}{c|}{\cellcolor[HTML]{E41A1C}}                         & \multicolumn{1}{c|}{\cellcolor[HTML]{E41A1C}}                         & \multicolumn{1}{c|}{\cellcolor[HTML]{E41A1C}}                        & \multicolumn{1}{c|}{\cellcolor[HTML]{E41A1C}}                         & \multicolumn{1}{c|}{\cellcolor[HTML]{E41A1C}}                         & \multicolumn{1}{c|}{\cellcolor[HTML]{E41A1C}}                         & \multicolumn{1}{c|}{\cellcolor[HTML]{E41A1C}}                         & \multicolumn{1}{c|}{\cellcolor[HTML]{E41A1C}}                          & \multicolumn{1}{c|}{\cellcolor[HTML]{E41A1C}}      & \multicolumn{1}{c|}{\cellcolor[HTML]{E41A1C}}      & \cellcolor[HTML]{E41A1C}      \\ \cline{1-3} \cline{5-5} \cline{7-21} 
\multicolumn{1}{|l|}{UE Radio } & \multicolumn{2}{l|}{Capabilities Tranfer} &     & \cellcolor[HTML]{377EB8}  &   & \multicolumn{1}{l|}{\cellcolor[HTML]{4DAF4A}} & \multicolumn{1}{l|}{\cellcolor[HTML]{4DAF4A}} & \multicolumn{1}{l|}{\cellcolor[HTML]{4DAF4A}} & \multicolumn{1}{l|}{\cellcolor[HTML]{4DAF4A}} & \multicolumn{1}{l|}{\cellcolor[HTML]{4DAF4A}} & \multicolumn{1}{l|}{\cellcolor[HTML]{4DAF4A}} & \multicolumn{1}{l|}{\cellcolor[HTML]{E41A1C}}  & \multicolumn{1}{l|}{\cellcolor[HTML]{E41A1C}}   & \multicolumn{1}{l|}{\cellcolor[HTML]{E41A1C}}  & \multicolumn{1}{l|}{\cellcolor[HTML]{E41A1C}}  & \multicolumn{1}{l|}{\cellcolor[HTML]{4DAF4A}}  & \multicolumn{1}{l|}{\cellcolor[HTML]{4DAF4A}} & \multicolumn{1}{c|}{-----}   & \multicolumn{1}{c|}{-----}    & -----   \\ 
\cline{1-3} \cline{5-5} \cline{7-21} 
\multicolumn{1}{|l|}{UE Network}& \multicolumn{2}{l|}{Security Capabilities}&     & \cellcolor[HTML]{377EB8}  &   & \multicolumn{1}{l|}{\cellcolor[HTML]{E41A1C}} & \multicolumn{1}{l|}{\cellcolor[HTML]{E41A1C}} & \multicolumn{1}{l|}{\cellcolor[HTML]{E41A1C}} & \multicolumn{1}{l|}{\cellcolor[HTML]{E41A1C}} & \multicolumn{1}{l|}{\cellcolor[HTML]{E41A1C}} & \multicolumn{1}{l|}{\cellcolor[HTML]{E41A1C}} & \multicolumn{1}{l|}{\cellcolor[HTML]{E41A1C}}  & \multicolumn{1}{l|}{\cellcolor[HTML]{E41A1C}}   & \multicolumn{1}{l|}{\cellcolor[HTML]{E41A1C}}  & \multicolumn{1}{l|}{\cellcolor[HTML]{E41A1C}}  & \multicolumn{1}{l|}{\cellcolor[HTML]{E41A1C}}  & \multicolumn{1}{l|}{\cellcolor[HTML]{E41A1C}} & \multicolumn{1}{c|}{-----}   & \multicolumn{1}{c|}{-----}    & -----   \\ 
\cline{1-3} \cline{5-5} \cline{7-21} 
\multicolumn{1}{|l|}{}   & \multicolumn{2}{l|}{NAS Signalling}  &   & \cellcolor[HTML]{9698ED} &  & \multicolumn{1}{l|}{\cellcolor[HTML]{E41A1C}} & \multicolumn{1}{l|}{\cellcolor[HTML]{E41A1C}} & \multicolumn{1}{l|}{\cellcolor[HTML]{E41A1C}} & \multicolumn{1}{l|}{\cellcolor[HTML]{E41A1C}} & \multicolumn{1}{l|}{\cellcolor[HTML]{E41A1C}} & \multicolumn{1}{l|}{\cellcolor[HTML]{E41A1C}} & \multicolumn{1}{l|}{\cellcolor[HTML]{E41A1C}} & \multicolumn{1}{l|}{\cellcolor[HTML]{E41A1C}} & \multicolumn{1}{l|}{\cellcolor[HTML]{E41A1C}} & \multicolumn{1}{l|}{\cellcolor[HTML]{E41A1C}} & \multicolumn{1}{l|}{\cellcolor[HTML]{E41A1C}} & \cellcolor[HTML]{E41A1C} & \multicolumn{1}{c|}{\cellcolor[HTML]{E41A1C}}      & \multicolumn{1}{c|}{\cellcolor[HTML]{E41A1C}}      & \cellcolor[HTML]{4DAF4A}      \\ \cline{2-3} \cline{5-5} \cline{7-21} 
\multicolumn{1}{|l|}{}   & \multicolumn{2}{l|}{RRC Signalling}  &                     & \cellcolor[HTML]{9698ED}                      &                     & \multicolumn{1}{l|}{\cellcolor[HTML]{E41A1C}} & \multicolumn{1}{l|}{\cellcolor[HTML]{E41A1C}} & \multicolumn{1}{l|}{\cellcolor[HTML]{E41A1C}} & \multicolumn{1}{l|}{\cellcolor[HTML]{E41A1C}} & \multicolumn{1}{l|}{\cellcolor[HTML]{E41A1C}} & \multicolumn{1}{l|}{\cellcolor[HTML]{E41A1C}} & \multicolumn{1}{l|}{\cellcolor[HTML]{E41A1C}} & \multicolumn{1}{l|}{\cellcolor[HTML]{E41A1C}} & \multicolumn{1}{l|}{\cellcolor[HTML]{E41A1C}} & \multicolumn{1}{l|}{\cellcolor[HTML]{E41A1C}} & \multicolumn{1}{l|}{\cellcolor[HTML]{E41A1C}} & \cellcolor[HTML]{E41A1C} & \multicolumn{1}{c|}{\cellcolor[HTML]{4DAF4A}}      & \multicolumn{1}{c|}{\cellcolor[HTML]{4DAF4A}}      & \cellcolor[HTML]{4DAF4A}      \\ 
\cline{2-3} \cline{5-5} \cline{7-21}
\multicolumn{1}{|l|}{\multirow{-3}{*}{\begin{tabular}[c]{@{}l@{}}Confidentiality \\ Protection\end{tabular}}} & \multicolumn{2}{l|}{User Data}                                           &                     & \cellcolor[HTML]{9698ED}                      &                     & \multicolumn{1}{l|}{\cellcolor[HTML]{4DAF4A}} & \multicolumn{1}{l|}{\cellcolor[HTML]{4DAF4A}} & \multicolumn{1}{l|}{\cellcolor[HTML]{4DAF4A}} & \multicolumn{1}{l|}{\cellcolor[HTML]{4DAF4A}} & \multicolumn{1}{l|}{\cellcolor[HTML]{4DAF4A}} & \multicolumn{1}{l|}{\cellcolor[HTML]{4DAF4A}} & \multicolumn{1}{l|}{\cellcolor[HTML]{4DAF4A}} & \multicolumn{1}{l|}{\cellcolor[HTML]{4DAF4A}} & \multicolumn{1}{l|}{\cellcolor[HTML]{4DAF4A}} & \multicolumn{1}{l|}{\cellcolor[HTML]{4DAF4A}} & \multicolumn{1}{l|}{\cellcolor[HTML]{E41A1C}} & \cellcolor[HTML]{4DAF4A} & \multicolumn{1}{c|}{\cellcolor[HTML]{4DAF4A}}      & \multicolumn{1}{c|}{\cellcolor[HTML]{E41A1C}}      & \cellcolor[HTML]{E41A1C}      \\ \cline{1-3} \cline{5-5} \cline{7-21} 
\multicolumn{1}{|l|}{}                                                                                        & \multicolumn{2}{l|}{NAS Signalling}                                      &                     & \cellcolor[HTML]{377EB8}                      &                     & \multicolumn{1}{l|}{\cellcolor[HTML]{E41A1C}} & \multicolumn{1}{l|}{\cellcolor[HTML]{E41A1C}} & \multicolumn{1}{l|}{\cellcolor[HTML]{E41A1C}} & \multicolumn{1}{l|}{\cellcolor[HTML]{E41A1C}} & \multicolumn{1}{l|}{\cellcolor[HTML]{E41A1C}} & \multicolumn{1}{l|}{\cellcolor[HTML]{E41A1C}} & \multicolumn{1}{l|}{\cellcolor[HTML]{E41A1C}} & \multicolumn{1}{l|}{\cellcolor[HTML]{E41A1C}} & \multicolumn{1}{l|}{\cellcolor[HTML]{E41A1C}} & \multicolumn{1}{l|}{\cellcolor[HTML]{E41A1C}} & \multicolumn{1}{l|}{\cellcolor[HTML]{E41A1C}} & \cellcolor[HTML]{E41A1C} & \multicolumn{1}{c|}{\cellcolor[HTML]{4DAF4A}}      & \multicolumn{1}{c|}{\cellcolor[HTML]{4DAF4A}}      & \cellcolor[HTML]{4DAF4A}      \\ \cline{2-3} \cline{5-5} \cline{7-21} 
\multicolumn{1}{|l|}{}                                                                                        & \multicolumn{2}{l|}{RRC Signalling}                                      &                     & \cellcolor[HTML]{377EB8}                      &                     & \multicolumn{1}{l|}{\cellcolor[HTML]{E41A1C}} & \multicolumn{1}{l|}{\cellcolor[HTML]{E41A1C}} & \multicolumn{1}{l|}{\cellcolor[HTML]{E41A1C}} & \multicolumn{1}{l|}{\cellcolor[HTML]{E41A1C}} & \multicolumn{1}{l|}{\cellcolor[HTML]{E41A1C}} & \multicolumn{1}{l|}{\cellcolor[HTML]{E41A1C}} & \multicolumn{1}{l|}{\cellcolor[HTML]{E41A1C}} & \multicolumn{1}{l|}{\cellcolor[HTML]{E41A1C}} & \multicolumn{1}{l|}{\cellcolor[HTML]{E41A1C}} & \multicolumn{1}{l|}{\cellcolor[HTML]{E41A1C}} & \multicolumn{1}{l|}{\cellcolor[HTML]{E41A1C}} & \cellcolor[HTML]{E41A1C} & \multicolumn{1}{c|}{\cellcolor[HTML]{4DAF4A}}      & \multicolumn{1}{c|}{\cellcolor[HTML]{4DAF4A}}      & \cellcolor[HTML]{4DAF4A}      \\ \cline{2-3} \cline{5-5} \cline{7-21} 
\multicolumn{1}{|l|}{\multirow{-3}{*}{Integrity Protection}}                                                  & \multicolumn{2}{l|}{User Data}                                           &                     & \cellcolor[HTML]{9698ED}                      &                     & \multicolumn{1}{l|}{\cellcolor[HTML]{E41A1C}} & \multicolumn{1}{l|}{\cellcolor[HTML]{E41A1C}} & \multicolumn{1}{l|}{\cellcolor[HTML]{E41A1C}} & \multicolumn{1}{l|}{\cellcolor[HTML]{E41A1C}} & \multicolumn{1}{l|}{\cellcolor[HTML]{E41A1C}} & \multicolumn{1}{l|}{\cellcolor[HTML]{E41A1C}} & \multicolumn{1}{l|}{\cellcolor[HTML]{E41A1C}} & \multicolumn{1}{l|}{\cellcolor[HTML]{E41A1C}} & \multicolumn{1}{l|}{\cellcolor[HTML]{E41A1C}} & \multicolumn{1}{l|}{\cellcolor[HTML]{E41A1C}} & \multicolumn{1}{l|}{\cellcolor[HTML]{E41A1C}} & \cellcolor[HTML]{E41A1C} & \multicolumn{1}{c|}{\cellcolor[HTML]{E41A1C}}      & \multicolumn{1}{c|}{\cellcolor[HTML]{E41A1C}}      & \cellcolor[HTML]{E41A1C}      \\ \cline{1-3} \cline{5-5} \cline{7-21} 
\multicolumn{1}{|l|}{Confidentiality Mechanisms}                                                              & \multicolumn{2}{l|}{Supported by UE}                                     &                     & \cellcolor[HTML]{377EB8}                      &                     & \multicolumn{1}{l|}{\cellcolor[HTML]{4DAF4A}} & \multicolumn{1}{l|}{\cellcolor[HTML]{4DAF4A}} & \multicolumn{1}{l|}{\cellcolor[HTML]{4DAF4A}} & \multicolumn{1}{l|}{\cellcolor[HTML]{4DAF4A}} & \multicolumn{1}{l|}{\cellcolor[HTML]{4DAF4A}} & \multicolumn{1}{l|}{\cellcolor[HTML]{4DAF4A}} & \multicolumn{1}{l|}{\cellcolor[HTML]{4DAF4A}} & \multicolumn{1}{l|}{\cellcolor[HTML]{4DAF4A}} & \multicolumn{1}{l|}{\cellcolor[HTML]{4DAF4A}} & \multicolumn{1}{l|}{\cellcolor[HTML]{4DAF4A}} & \multicolumn{1}{l|}{\cellcolor[HTML]{4DAF4A}} & \cellcolor[HTML]{4DAF4A} & \multicolumn{1}{c|}{-----}                         & \multicolumn{1}{c|}{-----}                         & -----                         \\ \cline{1-3} \cline{5-5} \cline{7-21} 
\multicolumn{1}{|l|}{Integrity Mechanisms}                                                                    & \multicolumn{2}{l|}{Supported by UE}                                     & \multirow{-16}{*}{} & \cellcolor[HTML]{377EB8}                      & \multirow{-16}{*}{} & \multicolumn{1}{l|}{\cellcolor[HTML]{4DAF4A}} & \multicolumn{1}{l|}{\cellcolor[HTML]{4DAF4A}} & \multicolumn{1}{l|}{\cellcolor[HTML]{4DAF4A}} & \multicolumn{1}{l|}{\cellcolor[HTML]{4DAF4A}} & \multicolumn{1}{l|}{\cellcolor[HTML]{4DAF4A}} & \multicolumn{1}{l|}{\cellcolor[HTML]{4DAF4A}} & \multicolumn{1}{l|}{\cellcolor[HTML]{4DAF4A}} & \multicolumn{1}{l|}{\cellcolor[HTML]{4DAF4A}} & \multicolumn{1}{l|}{\cellcolor[HTML]{4DAF4A}} & \multicolumn{1}{l|}{\cellcolor[HTML]{4DAF4A}} & \multicolumn{1}{l|}{\cellcolor[HTML]{4DAF4A}} & \cellcolor[HTML]{4DAF4A} & \multicolumn{1}{c|}{-----}                         & \multicolumn{1}{c|}{-----}                         & -----                         \\ \cline{1-3} \cline{5-5} \cline{7-21} 
\end{tabular}}
\vspace{0.5 mm}
    \begin{center}
        \fbox{\begin{tabular}{ll | ll | ll | ll }
                    \textcolor{Mandatory5G}{$\blacksquare$} & 5G SA Mandatory (TS 33.501 \cite{ts33501})
                &   \textcolor{Optional5G}{$\blacksquare$} & 5G SA Optional (TS 33.501 \cite{ts33501})
                &   \textcolor{In5gStandard}{$\blacksquare$} & 5G Compliant
                &   \textcolor{NoSecurity}{$\blacksquare$} & No 5G Compliant
        \end{tabular}}
        \vspace{2mm}
        \caption{5G Security mechanisms availability on current network deployments}
        \label{table:reality_check_results}
    \end{center}    
    \vspace{-6mm}
\end{table*}

\subsection{Data Evaluation Methodology}
\label{subsec:data_eval_methodology}

Traces extracted from the communication process contains all the information required for the evaluation of 5G networks security features introduced in Section~\ref{sec:security_in_5gnr}. We look into the RRC and NAS messages to identify the status of the security enhancements.

\textbf{Deployment type identification}. The first step in the evaluation process is to identify the type of deployment at which the user was connecting to. The incremental approach followed by operators towards the deployment of 5G networks results in two different types of deployments: i) 5G NSA and 5G SA (see Sec.~\ref{sec:background}).
The identification between NSA and SA will be performed by using the Information Elements (IEs) carried by the MIB. More specifically, in 5G SA deployments the gNB will include \textit{pdcch-ConfigSIB1}, \textit{ssb-SubcarrierOffset} or \textit{dmrs-TypeA-Position} IEs, which will not be present on a 5G NSA deployment.

\textbf{Authentication procedure}. The evaluation of the authentication procedure will be performed after the RRC connection establishment. Apart from the different messages exchange from other authentication procedures, there are other indicators within the messages that allow the proper identification of 5G AKA. For example, after the \textit{RRCSetupComplete} message, the UE sends a \textit{NAS RegistrationRequest} initiating the authentication procedure and hence, disclosing the underlying authentication procedure (e.g. "5GS registration type" field, 5G-GUTI as \textit{TypeOfIdentity}, or the inclusion of the 5G-TMSI). 

\textbf{Privacy and Anonymity}. Privacy and anonymity of terminals depend on whether UE identity is accessible by third-party observers or not. There are two types of parameters devoted to identify UEs within the authentication process: i) the permanent subscriber identifiers, which must be securely transmitted; and ii) temporal subscriber identifiers, which must be periodically updated in order to avoid their correlation with UEs. The permanent subscriber identifier can be found in the \textit{NAS RegistrationRequest} (within the \textit{5GS Mobile Identity} IE), when the UE starts a registration procedure or in the \textit{NAS IdentityResponse} after receiving a \textit{NAS IdentityRequest} from the network.
Then, we focus on measuring the refresh rate of the temporal identifiers (5G-GUTI and 5G-TMSI). Their values must be updated after each registration procedure within \textit{NAS Registration Accept} message and after \textit{NAS Service Request} in the subsequent \textit{RRC Connection Request} message where a new value for 5G-TMSI shall be assigned by the gNB.
 
In order to assess the implementation of the new 5G security features, we will check if the security of the permanent identifiers and the refresh period was applied to the temporal subscriber identifiers by checking their value within the aforementioned messages.

\textbf{Confidentiality and Integrity}. To assess confidentiality and integrity in the Control Plane we need to look into the RRC and NAS \textit{SecurityModeCommand} messages, where the algorithms to provide protection are selected and activated. For the UP, we have first located the \textit{NAS Service Request} and the subsequent \textit{RRC SecurityModeCommand} messages, which activate the Data Radio Bearer (DRB) and the algorithms. However, the UP security is established with the \textit{RRCReconfiguration} message, which carries information about the algorithms used for the service to provide integrity and confidentiality protection per DRB.

\textbf{UE Supported Capabilities}. 
UE network capabilities are always sent in the \textit{NAS Registration Request} message within the UE network capability and UE additional security capability fields. In these both fields, all the security algorithms supported by the UE regarding each mobile technology are sent to the base station. In the case of NSA deployments, this information is carried by the \textit{UECapabilityInformation} message sent by the UE.

\textbf{UE Radio Capabilities Transfer}. 
UE radio capabilities are sent in the \textit{RRC UE Capability Information} message. Following the registration procedure time events in the traces, we verify that in some networks this message is sent before the \textit{RRC Security Mode Command} without confidentiality or integrity protection.

\section{Evaluation}
\label{sec:evaluation}

\subsection{Reality Check, is current 5G Really Improving Security?}
\label{subsec:reality_check}
The results of the 
analysis following the methodology introduced in Section~\ref{subsec:data_eval_methodology} are summarised in Table~\ref{table:reality_check_results}. Each row of the table represents the different security features under study, and being the columns, the standard view of each feature, the results obtained for two different operators and the results obtained in \cite{nie2022measuring} respectively.

The first result to highlight is the complete absence of 5G SA deployments. Both operators are offering 5G coverage by means of 5G NSA deployments which essentially rely on existing 4G infrastructures. Hence, there is no enhancement on the Authentication and Key Agreement process.

\textbf{Ciphering of Permanent Identifiers}:
We have checked that no concealment of permanent identifiers has been done by capturing the permanent IMSI and IMEI values which are sent without protection within the \textit{NAS Identity Response} message.

\textbf{Temporary Identifier and GUTI Refresh:} 
We have verified along the different traces, after receiving the \textit{NAS Attach Accept} and \textit{RRC Connection Request} messages, the freshness of \textit{m-TMSI} value within GUTI. \textit{m-TMSI} shall change its value after these messages, however, only during the Registration procedure the temporary identifier is updated.

\textbf{Confidentiality and Integrity:}
In terms of confidentiality, on the one hand the \textit{nr- RadioBearerConfig-r15} IE to establish the 
DRB points to the \textit{NEA2} algorithm in all the traces except for Operator B in the city of Tarragona. This algorithm indicates that User Data confidentiality is effectively met even if the standard marks it as optional. In contrast, Tarragona does not accomplish confidentiality protection of user data due to the lack of a 5G DRB in this area. On the other hand, confidentiality protection for the initial NAS and RRC messages is not yet implemented using 5G NEA algorithms.

In contrast to confidentiality in data transmissions, integrity is a mandatory feature for signalling messages. Nevertheless, the configured data and signalling radio bearers do not show any of the mandatory algorithms in the \textit{IntegrityProtAlgorithm} field within the IEs. Instead, they use algorithms from previous generations (i.e., eia2) which do not provide the required security level.

\textbf{UE Network security capabilities}: Moreover, we have verified the supported algorithms in the UE by checking the UE security capabilities within \textit{NAS Attach Request} message. Despite only using 5G NEA algorithm for securing the UP, the UE supports both 5G NEA and NIA plus legacy 4G and 3G algorithms.

\textbf{UE Radio capabilities}: We found that only in Operator B, four access networks are sending the radio capabilities before initialising the security environment for the CP messages.

Although there is a clear trend in the reported results, note that there might be other operators/deployments (not covered in this measurement campaign) exhibiting better security results if they are more advanced in their migration path from 4G to 5G.

\subsection{Effective Attacks on current 5G Deployments}
\label{subsec:effective_attacks}

\textbf{Subscriber credentials (identity attacks)}:
Since none of the studied networks implement concealment of the permanent identifier, the legacy \textit{IMSI} catching attacks can still be deployed \cite{IMSI_catching} as well as more sophisticated attacks that exploit subscriber credentials \textit{leakability} \cite{chlosta20215g} \cite{sec5G_Nori}.
Moreover, temporary identifiers can be found in all captures (updated every time the Registration Procedure is performed), enabling identity mapping and tracking attacks by correlating temporary identifiers with UEs.

\textbf{Authentication vulnerabilities (activity monitoring)}: 
Our previous section revealed the complete absence of 5G-AKA protocol and hence, the presence of UEs and their consumed mobile services can be inferred. Authors in \cite{borgaonkar2019new} propose novel privacy attacks against all variants of \textit{AKA} protocol which also affect the studied scenarios.

\textbf{UP Confidentiality and Integrity}: 
As highlighted in the evaluation section, confidentiality protection is enabled in most of the studied deployments while integrity protection is completely missing. This absence allows an adversary to perform data manipulation, identity mapping and impersonation attacks (i.e., \textit{MitM} attacks) even if confidentiality is active \cite{sec4G_BreakingLTE_L2,rupprecht2020imp4gt,kohls2019lost}.

\textbf{UE Radio Capabilities}: 
Transmitting radio capabilities information before the CP security activation (\textit{Security Mode Command} message) could lead to Identification, Binding Down and Battery Drain attacks \cite{shaik2019new}. Given the obtained results, most of the deployments enclosed by Operator B are susceptible to these attacks.

The implementation of active data collection methodologies (e.g. \cite{chlosta2019lte}, \cite{chlosta20215g}) would enrich the obtained results, allowing an in-depth analysis of the security features not only from a network subscriber perspective but from the view of an active attacker willing to exploit the available vulnerabilities.

\begin{table}[t!]
    \centering
    \resizebox{\columnwidth}{!}{\begin{tabular}{c c c c c}
    \toprule 
    \textbf{Security} & \textbf{5G NSA} & \textbf{Threats} & \textbf{5G SA}\\
    \textbf{Field} & Vulnerability & \textbf{and Attacks} & Enhancement  \\
      \midrule 
                            & No concealment of     & \cite{chlosta20215g}, \cite{sec5G_Nori}, & Concealment of \\
        \textbf{Subscriber} & permanent identifiers & \cite{IMSI_catching} & SUPI, the SUCI \\
        \textbf{Credentials}& No specific policies  & \cite{sec4G_BreakingLTE_L2}, \cite{shaik2015practical} & GUTI reallocation after  \\
                            & for GUTI reallocation &  & Registration and Service Request \\
        \midrule
    \textbf{Authentication} & Lack of randomness and & \cite{borgaonkar2019new} & \_\_\_ \\
    \textbf{procedure} & the use of XOR in AUTS &                          &  \\
       \midrule 
    \textbf{UP}              & Optional & \cite{sec4G_BreakingLTE_L2}, \cite{shaik2015practical} & UE and gNB \\
    \textbf{Confidentiality} & Support &  & Mandatory Support \\
       \midrule
    \textbf{UP}        & Optional & \cite{sec4G_BreakingLTE_L2}, \cite{rupprecht2020imp4gt}, & UE and gNB \\
    \textbf{Integrity} & Support & \cite{kohls2019lost} & Mandatory Support \\
        \midrule
    \textbf{UE} & No security transfer & \cite{shaik2019new} & CP Security before  \\
    \textbf{Capabilities} & of UE Capabilities   &                     & transfer of Capabilities \\
    \bottomrule 
    \end{tabular}}
    \caption{Overview of vulnerabilities and attacks}
    \label{tab:Attacks_overview}
    \vspace{-11mm}
\end{table}

\section{Conclusions}
\label{sec:conclusions}
5G networks are expected to significantly improve the security of mobile users,  thanks to the newly introduced mandatory features which address identified 4G vulnerabilities. In this paper, we analysed the progress of current 5G European 
commercial networks deployments with respect to the expected security features. In order to do so, we collected a dataset comprising 5G measurements from two different operators in Spain, six different cities and both urban and suburban scenarios. The two major network operators considered in our study operate in 70\% of the European countries and, due to economies of scale, our  results can be reasonably expected to be applicable to other European countries served by the same operators. 
Our results show that current 5G network deployments miss expectations on i) providing improved privacy and anonymity to subscriber identifiers (transmitting them in clear text), ii) refreshing often enough temporal subscriber identifiers ( facilitating subscriber identification and tracking), iii) additional confidentiality protection (inheriting security vulnerabilities from previous generations) and iv) UE radio capabilities are sometimes transferred without protection (enabling bidding down and battery drain attacks).

As already reported in ~\cite{GSMA-5G}, we are in the midst of the 4G to 5G migration, expected to be mature by 2025. Thus, as we get closer to this date, we expect operators to increasingly deploy 5G security features accordingly, covering the gaps identified by our work.

\section{Acknowledgments}
This work has been supported by the Spanish Ministry of Economic Affairs and Digital Transformation and the European Union – NextGeneration EU, in the framework of the Recovery Plan, Transformation and Resilience (PRTR) (Call UNICO I+D 5G 2021, ref. number TSI-063000-2021-6-Open6G), and by the CERCA Programme from the Generalitat de Catalunya.

\bibliographystyle{ACM-Reference-Format}
\bibliography{references}

\end{document}